\documentclass[prl,twocolumn,showpacs,preprintnumbers,amsmath,amssymb]{revtex4}
\usepackage{graphicx}

\newcommand{\AP}{\mathrm{AP}}
\newcommand{\PP}{\mathrm{P}}
\newcommand{\LSMO}{La$_{2/3}$Sr$_{1/3}$MnO}
\newcommand{\STO}{SrTiO$_3$}
\newcommand{\ur}{$\uparrow$}
\newcommand{\dr}{$\downarrow$}

\pacs{75.70.Cn,73.20.-r,85.75.-d}

\begin{document}

\title{Half-metallic ferromagnets for magnetic tunnel junctions}

\author{Phivos Mavropoulos}\email{Ph.Mavropoulos@fz-juelich.de}        
\author{Marjana Le\v{z}ai\'{c}}
\author{Stefan Bl\"ugel}

\affiliation{Institut f\"ur Festk\"orperforschung, Forschungszentrum
  J\"ulich, D-52425 J\"ulich, Germany}

\date{\today}

\begin{abstract}
Using theoretical arguments, we show that, in order to exploit
half-metallic ferromagnets in tunneling magnetoresistance (TMR)
junctions, it is crucial to eliminate interface states at the Fermi
level within the half-metallic gap; contrary to this, no such problem
arises in giant magnetoresistance elements. Moreover, based on an {\it
a priori} understanding of the electronic structure, we propose an
antiferromagnetically coupled TMR element, in which interface states
are eliminated, as a paradigm of materials design from first
principles. Our conclusions are supported by {\it ab-initio}
calculations.
\end{abstract}

\maketitle


Half-metallic ferromagnets are ferromagnetic materials showing, in the
ideal case, 100\% spin polarization at the Fermi level $E_F$, due to a
metallic density of states in one spin direction (usually majority
spin) combined with a band gap in the other spin direction (usually
minority spin). First discovered by {\it ab-initio} calculations by
de~Groot {\it et al.}~\cite{deGroot83}, these materials have drawn
strong attention because of their potential applications in the field
of spintronics.  In principle, half-metallic ferromagnets are ideal
spin injectors and detectors, because under moderate voltage they can
carry current in only one spin direction. Therefore, they also
constitute ideal components for Giant Magnetoresistant (GMR) and
Tunneling Magnetoresistant (TMR) devices, with two half-metallic leads
sandwiching a nonmagnetic normal metal spacer (in GMR) or a
semiconductor or insulator spacer (in TMR). There is, for instance,
the experimental result of Bowen and collaborators \cite{Bowen03} who
obtained a TMR ratio (relative change of resistance upon change of the
magnetization alignment of the leads) higher than 1800\% in a \LSMO
/\STO/\LSMO\ junction; this extreme value was attributed to the
half-metallicity of \LSMO. Motivated by such findings, we set forth to
gain theoretical understanding of the conditions under which
half-metals can be fully exploited in TMR devices.

The purpose of this article is twofold. First, we demonstrate by
theoretical arguments on the electronic structure that it is much
easier to exploit the half-metallic property in a GMR element than in
a TMR one. We explain the implications caused by interface states in
TMR elements, and we suggest cases of improved TMR elements without
interface states. Then, we propose an antiferromagnetically coupled
TMR element (to serve as a magnetic field sensor) based on an {\it a
priori} understanding of the exchange interactions in such systems, as
a paradigm of materials design from first principles.


The idea of using half-metals in GMR and TMR junctions seems
simple. In a parallel (P) alignment of the magnetic moments of the
half-metallic leads sandwiching the spacer, \emph{some} current will
pass, either by metallic conduction (in GMR) or by tunneling (in TMR)
of majority spin electrons. On orienting the moments of the leads in
an antiparallel (AP) way, for one spin channel no current can enter
the junction (due to the minority-spin gap of the one lead), while in
the other spin direction no current can exit the junction (due to the
minority-spin gap of the other lead); thus \emph{no} current can
pass. Hence this is an ideal spin-controlled switch.

\begin{figure}
\begin{center}
\includegraphics*[angle=270,width=0.95\linewidth]{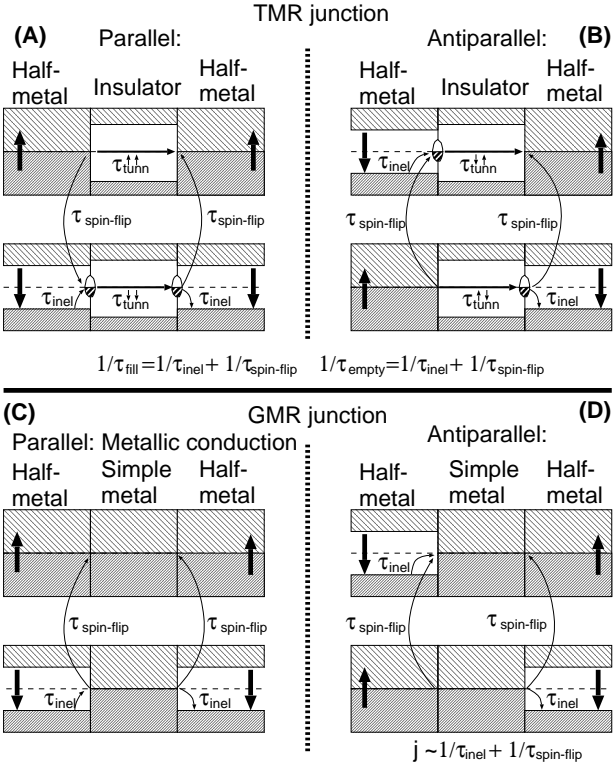}
\caption{Schematic band profile in TMR (A and B) and GMR (C and D)
  junctions using half-metallic leads.  In the middle of the gaps,
  $E_F$ is shown as a dashed line. Filled bands are shown as dark
  shaded regions, empty bands are lightly shaded; unshaded regions
  correspond to the band gaps.  In A and C, the parallel magnetic
  alignment of the leads is shown for both spin directions; in B and D
  the antiparallel one. In the TMR case (A and B) there is the
  possibility of interface states within the half-metallic gap, at
  $E_F$. Electrons can enter the interface states on the left at a
  rate $1/\tau_{\mathrm{fill}}$ and sequentially tunnel at a rate
  $1/\tau_{\mathrm{tunn}}$ (and similarly exit the interface states on
  the right). The time $\tau_{\mathrm{fill}}$ depends on inelastic
  processes and on the spin-flip rate, which can be much faster than
  the tunneling rate. Then $\tau_{\mathrm{tunn}}$ determines the
  current, and the half-metallic property is irrelevant. In GMR (C and
  D) no such problem appears, since there is metallic conduction in
  the parallel magnetic alignment.\label{fig:junct1}}
\end{center}
\end{figure}

However, in TMR junctions a difficulty arises in the presence of
interface states around $E_F$ in the half-metallic gap at the
metal-insulator contact. Consider, for instance, the TMR junction in
Fig.~\ref{fig:junct1} (A and B), where the bands are shown
schematically along the junction. Panel~A shows the band alignment for
both spin directions for a P alignment of the magnetzation of the
half-metallic leads, while panel~B shows the same for an AP alignment.
At the interfaces, for the minority-spin direction, possible localized
interface states are shown. If they exist, it is inevitable that they
are coupled to the bulk states of the half-metal, and thus they can be
important for the transport properties and for the TMR ratio of the
junction, as we will now discuss.

If interface states are present, they contribute to the tunneling
current $j$. The current is controlled by two sequential processes:
(i) by the tunneling itself, characterized by a rate
$1/\tau_{\mathrm{tunn}}$, and (ii) by re-filling the interface states
after an electron has tunneled out of them, with a characteristic rate
of $1/\tau_{\mathrm{fill}}$ (or by emptying these states after an
electron has tunneled into them, with a rate of
$1/\tau_{\mathrm{empty}}$) (otherwise they are blocked by the Pauli
principle or by Coulomb blockade effects).  Since these processes take
place sequentially, the characteristic times $\tau_{\mathrm{tunn}}$
and $\tau_{\mathrm{fill(empty)}}$ must be additive.  Then, in the AP
alignment the current $j_{\AP}$ has a non-zero value, and expression
for $j_{\AP}$ has the form (see also Fig.~\ref{fig:junct1}B)
\begin{equation}
j_{\AP} \sim \frac{1}{\tau_{\mathrm{fill}}+\tau_{\mathrm{tunn}}^{\uparrow\downarrow}}
+ \frac{1}{\tau_{\mathrm{tunn}}^{\downarrow\uparrow}+ \tau_{\mathrm{empty}}}.
\label{eq:1}
\end{equation}
The first term refers to filling a spin-down interface state at the
left lead in Fig.~\ref{fig:junct1}B (up) and tunneling to the right
lead, while the second term refers to tunneling from the spin-up
continuum of the left lead in Fig.~\ref{fig:junct1}B (down) into the
interface state of the right lead, and then emptying it. We
distinguish among four different tunneling times, for the four
different cases of tunneling between majority and minority states as
shown in Fig.~\ref{fig:junct1}~A and~B. We we name these
$\tau_{\mathrm{tunn}}^{\uparrow\uparrow}$,
$\tau_{\mathrm{tunn}}^{\downarrow\downarrow}$,
$\tau_{\mathrm{tunn}}^{\downarrow\uparrow}$, and
$\tau_{\mathrm{tunn}}^{\uparrow\downarrow}$.  Evidently the slower of
the two processes (i) and (ii) determines the current. If, in
comparison to the slow tunneling rate, the states are immediately
refilled (or emptied) after a tunneling event (we will argue below
that this is expected), then
$\tau_{\mathrm{fill(empty)}}\ll\tau_{\mathrm{tunn}}$ and $j_{\AP}$ is
determined by the tunneling rate alone, irrespectively of the
half-metallic band gap. Similar considerations hold for the
minority-spin current in the P case.

What determines the coupling of the interface states with the bulk and
thus the characteristic times $\tau_{\mathrm{fill}}$ and
$\tau_{\mathrm{empty}}$? On one hand, there are inelastic processes
contributing with a rate $1/\tau_{\mathrm{inel}}$. These can be of
thermal nature or quantum fluctuations (scattering of electrons with
phonons, magnons, other electrons {\it etc.}).  Usually inelastic
processes are slow at low temperatures, but if the Fermi level is in
the proximity of the band edges, rather than in mid-gap, they can be
of significance. More importantly, there is always some spin-orbit
coupling present. Therefore even in the bulk of the half-metal the
polarization at $E_F$, $P(E_F)$, is always lower than the ideal 100\%;
{\it e.g.}, $P(E_F)\simeq 99\%$ for NiMnSb \cite{Mavropoulos04}, and
the value decreases when the material is composed by heavier elements
or when $E_F$ is touching the band edges ({\it e.g.}, $P(E_F)\simeq
67\%$ for PtMnSb) \cite{Mavropoulos04}. Spin-orbit coupling will
contribute to filling or emptying the interface states with a rate of
$1/\tau_{\mathrm{spin-flip}}$. This acts in parallel with the
inelastic processes, and thus
\begin{equation}
1/{\tau_{\mathrm{fill(empty)}}} = 1/{\tau_{\mathrm{inel}}}
+ 1/{\tau_{\mathrm{spin-flip}}}
\label{eq:2}
\end{equation}
Additional factors can come into this equation in the presence of
defects or impurities which reduce $P(E_F)$ by introducing gap states.
For majority electrons we do not discuss the interface states
separately than the bulk states, since they are irrelevant for the
half-metallic property; their effect is included in
$\tau_{\mathrm{tunn}}$.

Although the rate $1/\tau_{\mathrm{fill(empty)}}$ in Eq.~(\ref{eq:2})
is low, we recall that tunneling can be a very slow process
($\tau_{\mathrm{tunn}}$ is long, growing exponentially with insulator
thickness and barrier height). Therefore, for thick or high insulating
barriers the interface states are immediately re-filled (or
re-emptied) after each tunneling event, and they act as a reservoir of
electrons. The fact that they are much weaker coupled to the bulk than
the majority-spin states does not help, since everything is determined
by the much slower tunneling time.  Assuming then that all tunneling
times $\tau_{\mathrm{tunn}}$ are long, Eqs.~(\ref{eq:1}) and
(\ref{eq:2}) lead to
\begin{equation}
j_{\AP} \sim 
\frac{1}{\tau_{\mathrm{tunn}}^{\downarrow\uparrow}}+
\frac{1}{\tau_{\mathrm{tunn}}^{\uparrow\downarrow}}
\mbox{\ \ and\ \ }
j_{\PP} \sim 
\frac{1}{\tau_{\mathrm{tunn}}^{\uparrow\uparrow}}+
\frac{1}{\tau_{\mathrm{tunn}}^{\downarrow\downarrow}}
\label{eq:3}
\end{equation}
This means, that the current depends only on the tunneling rates for
the two spin directions and not at all on the half-metallic property
of the lead.

The tunneling rates themselves depend on numerous factors: the
insulating barrier thickness, the details of the interface structure,
the presence of interface disorder, the symmetry character of the
interface states, the presence of defects in the insulating spacer,
etc.. Particularly important is the spin polarization $P(E_F)$ at the
interface~\cite{Kammerer04}. This, in the absence of interface states,
is approximately the same as in the bulk of the half-metal, but in
their presence it can have a completely different value and can even
be reversed \cite{Galanakis04}. The influence of these factors is in
general different on each of the four tunneling times
$\tau_{\mathrm{tunn}}^{\downarrow\uparrow}$,
$\tau_{\mathrm{tunn}}^{\uparrow\downarrow}$,
$\tau_{\mathrm{tunn}}^{\uparrow\uparrow}$, and
$\tau_{\mathrm{tunn}}^{\downarrow\downarrow}$, since the nature of the
states involved is different. Thus \emph{some} TMR ratio can appear,
but no extraordinary effect can be guaranteed by the half-metallic
property, unless one can eliminate the interface states. We note that
if these are eliminated, there is still a low rate of incoming
minority-spin states from the bulk to the interface, because of the
spin-orbit coupling. This rate, however, is very low (determined by
the high polarization $P(E_F)$).

In GMR junctions, on the other hand, the interface states play no
significant role, as demonstrated schematically in
Fig.~\ref{fig:junct1}~C and~D. In the P case the conduction is
metallic, while in the AP case it is confined at most to the value of
the spin polarization at $E_F$ in the bulk of the half-metallic leads
(plus inelastic effects); if this is determined by the spin-orbit
coupling, it should lead to an effect of the order of 1\%. This means
that the half-metallic property is fully exploited in the case of GMR,
in contrast to TMR.

At this point we conclude that, in order to exploit the half-metallic
property in TMR junctions, we must find half-metal / insulator
interfaces without interface states; and to this we now turn.

The most studied half-metallic ferromagnets are probably Heusler
alloys. The bulk band structure and the origin of the gap are well
understood \cite{Galanakis02a}, and so are their surface
\cite{Galanakis02c} and interface \cite{Galanakis04}
properties. Unfortunately, calculations of Heusler alloy /
semiconductor interfaces are conclusive on the appearance of interface
states at $E_F$ in almost all cases. Thus, our previous analysis rules
out Heusler alloys as good candidates for TMR junctions.

On the other hand, the class of half-metallic zinc-blende pnictides
and chalcogenides shows no interface states at $E_F$ when brought in
contact with zinc-blende (zb) semiconductors
\cite{Mavropoulos04b}. The reason is that, here, the gap originates
from a hybridization and repulsion of the transition-metal $d$ states
with the $p$ states of the $sp$ anion. This continues coherently at
the interface between the $sp$ anion and the cation of the
semiconductor. No unsaturated bonds are left to produce spurious
interface states. Such compounds (in particular CrAs \cite{Akinaga00},
CrSb \cite{Zhao01}, and small islands of MnAs \cite{Ono02}) have
already been experimentally realized by molecular beam epitaxy, and
show Curie points well above room temperature. Also multilayers
of CrAs and CrSb with GaAs have been made \cite{Mizuguchi02,Zhao04}.
Therefore, we consider this class of compounds well suited for TMR
junctions.


In magnetic field sensor applications of GMR and TMR, it is desirable
that the leads of the junction are coupled magnetically AP in the
ground state; then, with the application of a magnetic field, the
leads are re-oriented in a P fashion, and the conductance
changes. Moreover, the energy difference $\Delta E$ between AP and P
should be small enough that the switching occurs at moderate
fields. In GMR, both the property of AP coupling and the coupling
energy can be tuned by changing the spacer thickness $d$, since
$\Delta E(d)$ follows a decaying, oscillating pattern
\cite{Bruno95}. In the case of TMR, increasing $d$ results in an
exponential, but not oscillating, decoupling of the two
leads. Therefore, we seek TMR systems where the AP coupling is
dictated by {\it a priori} known physical properties, while $|\Delta
E|$ can be tuned {\it a posteriori} by changing the insulator spacer
thickness.  Again this can be achieved by using half-metallic zb
compounds.

The magnetic coupling in such zb compounds is well understood
\cite{Shirai01,Galanakis03}. The origin of ferromagnetism is mainly
the broadening of the majority $p$-$d$ hybrid band, whenever it is
partly occupied (the double exchange mechanism). This is the case,
{\it e.g.}, for CrAs, MnAs, and CrTe. On the contrary, FeAs and MnTe
have one electron too much: the majority $p$-$d$ band is fully
occupied, so that no energy is gained by its broadening, and the
antiferromagnetic susceptibility prevails.

The zinc-blende structure, along the $\langle 001\rangle$ direction,
can be viewed as an epitaxial structure of chemically alternating
atomic layers. For example, CrTe has alternating layers of Cr and Te
in the form $\cdots \mathrm{CrTeCrTe}\cdots$.  We interrupt this
succession by introducing semiconducting CdTe layers which decouple
two CrTe leads. The structure will have the form $\cdots
\mathrm{CrTeCr\underline{TeCdTe}CrTeCrTe}\cdots$. This structure is
still ferromagnetic and half-metallic with no interface states at
$E_F$ (we verified this by {\it ab-initio} calculations). But now we
introduce one layer of Mn at the CrTe/CdTe interface to cause an AP
coupling of the leads. The layer-by-layer structure will be
$\cdots\mathrm{Cr}\mathrm{Te}\mathrm{Cr}\mathrm{Te}\mathrm{\fbox{Mn}}\mathrm{\underline{TeCdTe}}\mathrm{\fbox{Mn}}\mathrm{Te}\mathrm{Cr}\mathrm{Te}\mathrm{Cr}\cdots$
The AP coupling is expected because of the Mn-Mn interaction, by the
same mechanism which brings MnTe to an antiferromagnetic \cite{Wei87}
state. The idea of this interface engineering is to introduce an
element with higher number of valence electrons at the interface (here
Mn in the place of Cr), so that the double exchange mechanism is not
any more present, because the bands are filled.

\begin{table}
\begin{tabular}{cccccccccccccc}
$\cdots$ & Cr & Te & Cr & Te & Mn &\underline{TeCdTe}& Mn & Te & Cr & Te & Cr&$\cdots$&\\
$\cdots$ &\ur &    &\ur &    &\dr &                  &\ur &    &\dr &    &\dr&$\cdots$&(AP)\\ 
$\cdots$ &\ur &    &\ur &    &\dr &                  &\dr &    &\ur &    &\ur&$\cdots$&(P)
\end{tabular}
\caption{The proposed half-metallic TMR element. The arrows indicate
  the calculated magnetic moment direction in each layer. The ground
  state is AP with the P state 15~meV higher. More CdTe layers will
  provide further decoupling.\label{tab:1}}
\end{table}

\begin{figure}
\begin{center}
\includegraphics*[width=0.8\linewidth]{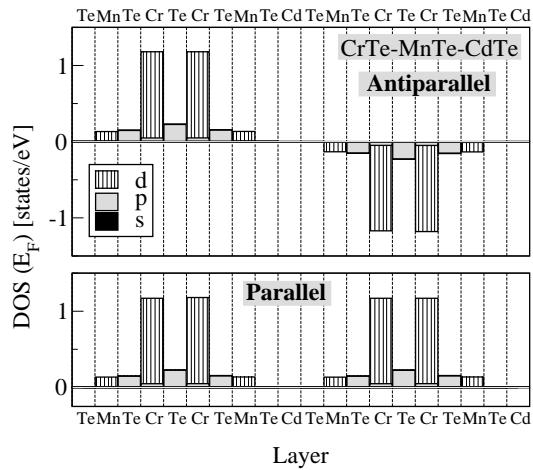}
\caption{Layer-resolved density of states (DOS) at $E_F$ for the
  junction shown in Table~\protect{\ref{tab:1}} in the ground state
  (top) (AP alignment) and also in the P alignment
  (bottom). (The symmetry resolution of the DOS is also given;
  negative DOS corresponds to spin-down electrons). Each lead by
  itself is half-metallic, and there are no minority-spin interface
  states at $E_F$. Thus, in the antiparallel case no current can
  pass. In the parallel case there can be tunneling of spin-up
  electrons.}
\label{fig:junct3}
\end{center}
\end{figure}

We verified these predictions by first-principles calculations. We
used the full-potential linearized augmented plane-wave method as
implemented in the {\tt FLEUR} code, within the generalized gradient
approximation of density-functional theory, using the CdTe lattice
constant. In Table~\ref{tab:1} we present the calculated geometry in
more detail. A supercell was used in the calculation, consisting of
two ``leads'', each having two Cr and two Mn layers (and corresponding
Te layers), separated by a CdTe layer for decoupling. Various possible
magnetic configurations were examined. In the ground state (AP in
Table~\ref{tab:1}), the leads are AP coupled, as expected. In
addition, the Mn atoms are antiferromagnetically coupled to the Cr
atoms.
The nice feature is that, in the ground state, each lead is by itself
half-metallic, so that the whole system is non-conducting. This is
evident also from Fig.~\ref{fig:junct3}, where the layer-resolved
density of states at $E_F$ is shown. Spin-down electrons are blocked
in the first half of the junction, whereas spin-up electrons are
blocked in the second part.  By applying an external magnetic field
the system switches to the P configuration with an energy cost of
15~meV (per CdTe slab). The P state is half-metallic throughout the
junction and conducting by tunneling of spin-up electrons,
as can be seen in Fig.~\ref{fig:junct3} (bottom).

The switching energy from the AP to the P state can be tuned by
introducing more CdTe layers. To show this, we compare the case without
a CdTe layer (the interface is then of the form Mn-Te-Mn), where
$\Delta E=124$~meV, to the case of Mn-CdTe-Mn ($\Delta E=15$~meV), and
then to the case of Mn-CdTeCdTe-Mn ($\Delta E=3.6$~meV). Each
additional CdTe layer lowers the energy difference by an order of
magnitude. One or two more CdTe layers should decouple the layers
sufficiently.


In summary, we have discussed the use of half-metallic ferromagnets in
TMR and GMR junctions. We concluded that, while in GMR junctions the
half-metallic property can be exploited fully, in TMR junctions the
same property does not help if there are interface states present at
$E_F$ within the half-metallic gap of the half-metal / insulator
interface (as is typical for Heusler alloys). The reason is that the
tunneling rate is slow compared to the spin-flip rate, whence
minority-spin interface states are efficiently coupled to the metallic
reservoir of the majority-spin states.

We have also proposed that such TMR elements can be made by using
half-metallic zinc-blende pnictides and chalcogenides in contact with
II-VI semiconductors, which show no interface states at $E_F$. In this
case we showed that, under certain conditions, an antiparallel magnetic
coupling of the leads is possible, blocking the electric current
completely. Such a device will show ideal magnetoresistance ratio and
can serve as an ideal magnetic field sensor.


We are grateful to Prof.~H.~Akinaga, Prof.~P.~H.~Dederichs, and Dr.~G.~Bihlmayer
for fruitful discussions.

\end{document}